\documentclass[namedreferences]{solarphysics}

\usepackage[hyperref,optionalrh,showbiblabels]{spr-sola-addons} 
\usepackage{graphicx}        
\usepackage{color}           
\usepackage{breakurl}        
\usepackage{amsmath}

\renewcommand{\vec}[1]{{\mathbfit #1}}

\newcommand{\aap}{    {\it Astron. Astrophys.}}

\newcommand{\apj}{    {\it Astrophys. J.}}
\newcommand{\apjl}{   {\it Astrophys. J. Lett.}}

\newcommand{\grl}{    {\it Geophys. Res. Lett.}}

\newcommand{\jgr}{    {\it J. Geophys. Res. (Space Phys.)}}

\newcommand{\solphys}{{\it Solar Phys.}}
 
\newcommand{\ssr}{    {\it Space Sci. Rev.}} 
\chardef\us=`\_

\usepackage{hyperref}
\hypersetup{
    colorlinks=true,
    linkcolor=red,
    citecolor= blue,
    filecolor=red,      
    urlcolor=blue,
}
\begin{document}

\begin{article}
\begin{opening}

\title{A new tool for predicting the solar cycle: Correlation between flux transport 
at the equator and the poles}
 
\author[addressref={aff1},corref, email={susanta@nao.cas.cn}]{\inits{Susanta~Kumar~Bisoi}\fnm{Susanta~Kumar~Bisoi}~\lnm{}\orcid{0000-0002-9448-1794}}
\author[addressref={aff2},email={jerry@prl.res.in}]{\inits{P.~Janardhan}\fnm{P.~Janardhan}~\lnm{}\orcid{0000-0003-2504-2576}}

\address[id=aff1]{Key Laboratory of Solar Activity, National Astronomical Observatories, Chinese Academy of Sciences, Beijing 100 100, People's Republic of China}
\address[id=aff2]{Physical Research Laboratory, Astronomy \& Astrophysics Division, Navrangpura, Ahmedabad 380 009, India}

\runningauthor{S.~K.~Bisoi et al.}
\runningtitle{Correlation between flux transport at the equator and the poles}

\begin{abstract}

The magnetic flux cancellation on the Sun plays a crucial role in determining 
the manner in which the net magnetic flux changes in every solar cycle, affecting the large 
scale evolution of the coronal magnetic fields and heliospheric environment.
We investigate, in this paper, the correlation between the solar magnetic 
flux cancelled at the equator and the solar magnetic flux transported to 
the poles by comparing the net amount of magnetic flux in the 
latitude belt 0${^{\circ}}$--5${^{\circ}}$ and 45${^{\circ}}$--60${^{\circ}}$, 
estimated using synoptic magnetograms from National Solar Observatory at Kitt Peak, 
during Solar Cycles 21\,--\,24. We find a good correlation between the 
net flux in the latitude band 0${^{\circ}}$--5${^{\circ}}$ and 55${^{\circ}}$--60${^{\circ}}$ 
for the Northern as well as for the Southern hemispheres.  However, we find a poor 
correlation if the net flux for the Northern and Southern hemispheres are 
considered together. In addition, we investigate the correlation between the net flux 
cancelled at the equator and the strength of solar polar field at 
cycle minimum, and find a good correlation between the two. 
We discuss the implication of the correlation of 
flux transported across the equator and to the poles 
that has an important bearing in the estimation of the residual 
polar cap field strength at the cycle minimum. This can be used a 
predictive tool for estimating the amplitude of subsequent cycles and we use this 
to estimate maximum smoothed sunspot numbers of 77$\pm$5 and 85$\pm$5 for 
the Northern and Southern hemispheres, respectively, for the upcoming Solar Cycle 25. 

\end{abstract}

\end{opening}

\section{Introduction}\label{sec:intro} 

Detailed studies of solar photospheric magnetic fields are crucial for 
understanding the origin and evolution of solar magnetic activity which impacts 
the manner in which global solar magnetic fields evolve and change. 
The long term evolution of global solar
magnetic fields is eventually important in the context of the 
heliospheric environment \citep{Sch06,SaJ19}, space weather activity 
changes in near-Earth space \citep{Sch06b} and in understanding 
the terrestrial magnetosphere and ionosphere \citep{Pul07,JaB15b,
IJB19}. Typically the origin of the magnetic activity of 
the Sun, well known as the solar cycle and varying with a period 
of $\sim$11 years, is attributed to a cyclic process, operated 
by an interior solar dynamo, which generates toroidal fields by 
shearing pre-existing poloidal fields and eventually regenerating
poloidal fields to complete the cyclic process \citep{Cha10}. 
The toroidal fields get amplified by differential rotation 
\citep{Par55a} and rise through the turbulent convection zone 
to form sunspot pairs or bipolar magnetic regions (BMRs) 
\citep{Par55b}. The BMRs emerge typically at low solar latitudes, 
around 35${^{o}}$ North and South at the start of the solar cycle
constituting sunspot groups of positive and negative polarity 
fluxes. However, due to the systematic tilt of BMRs with respect 
to the equator \citep{HaE19}, caused by the action of the Coriolis 
force on the buoyantly rising toroidal magnetic flux tubes, one 
of the spots in BMRs leads the other in both hemispheres, commonly known as leading 
polarity flux. While the other sunspot is known as following polarity 
flux. These opposite sunspot polarity fluxes generally cancel among 
themselves and decay according to the Babcock-Leighton mechanism \citep{Bab61,Lei69}.  

The remnant leading and following polarity fluxes are transported, 
respectively, towards the equator and the poles due to diffusion 
and advection \citep{WNS89}. The leading polarity fluxes cancel 
those from the opposite solar hemisphere, at the equator, while 
the following polarity fluxes cancel the existing, opposite polarity
polar cap fields and impart a new polarity.  The polarity of the 
polar cap field thereby reverses, creating polar cap fields 
with opposite polarity. This well known process is referred to as 
polar field reversal, and with the exception of cycle 24 when the 
field reversal was extremely unusual and asymmetric \citep{GoY16,JaF18}, 
takes place at each solar maximum \citep{Bab61}. One can therefore 
expect that there should be a correlation between the net amount 
of flux cancelled at the equator and the net amount of flux transported 
to the poles. We thus, in this study, investigate the relationship
between the net flux variations across the equator and the poles 
during the past four solar cycles, {\it{viz.}} Cycles 21--24.

The polar field strength at the end of a cycle is crucial for 
determining the strength of toroidal fields in the subsequent 
cycle and, in turn, the amplitude of the next sunspot cycle \citep{JiW18}. In fact, 
the strength of polar field at the cycle minimum shows a 
good empirical correlation with the amplitude of the next sunspot 
cycle \citep{SPe93,Sch05}, so it has been used as a proxy to estimate 
the amplitude of the subsequent cycle \citep{ScS78,SFS98,SCK05,CCJ07,JaB15a}. Generally, 
surface flux transport simulations \citep{WNS89,She05,Mye12,IiH17,UHa18,JiW18,BNa18} 
have been used to estimate the strength of the polar field at the 
minimum of a cycle.  However, the possible correlation, if there is any, 
between the net flux strength across the equator and at the poles, can 
also be used to estimate the polar field strength at the cycle 
minimum. We thus discuss, in this study, the role of this correlation 
for estimating the amplitude of the upcoming Sunspot Cycle 25. 
\subsection{Magnetic Butterfly Diagram}
  \label{sec:mag-field}
\begin{figure*}[ht]
\vspace{7.65cm}
\center
\includegraphics{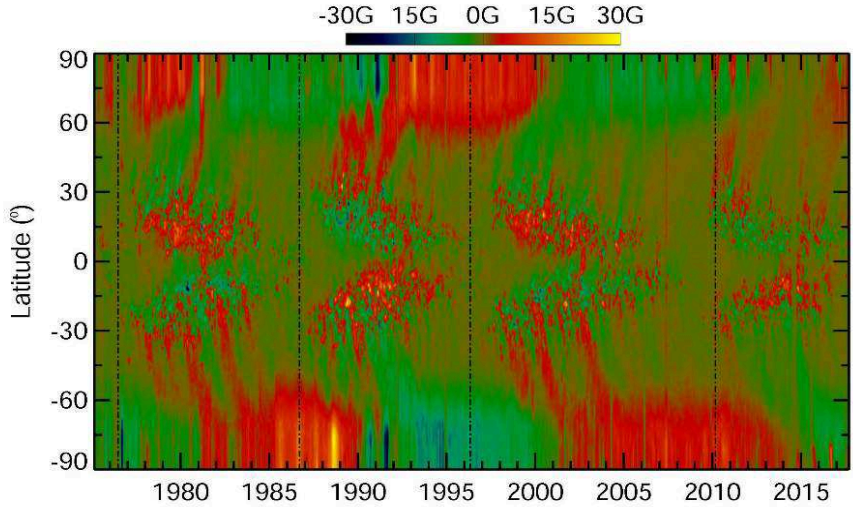}
\caption{A magnetic butterfly diagram shown,  for the period 
1975.14\,--\,2018.00, covering Solar Cycles 21\,--\,24. The magnetic flux 
has been limited to $\pm$30 Gauss so as to have better contrast and positive and 
negative polarities are shown in red and green respectively.   
The vertical dotted lines are drawn at solar minima of 
Cycles 20--23.}
\label{butter} 
\end{figure*}

The observational evidence of a correlation between the flux 
transport at the equator and the poles can be seen from a magnetic 
butterfly diagram. Figure \ref{butter} shows a magnetic butterfly 
diagram.  The diagram is basically a latitudinal map in time 
depicting the distribution of photospheric magnetic fields for the 
period from February 1975 (1975.12)\,--\,December 2017 (2018.00) 
or CR1625\,--\,CR2197 covering Solar Cycles 21\,--\,24.  A 
magnetic butterfly diagram is generated using photospheric magnetic 
field values obtained from Carrington Rotation (CR) averaged 
synoptic magnetograms.  A full description about the synoptic 
maps used is given in the following section and also can 
be found in \cite{JBG10,JaB15a,JaF18}. The procedure 
for generating the butterfly diagram is described briefly below, 
and more details can be found in \cite{BiJ14} and \cite{JaF18}.

The butterfly diagram depicts the evolution of sunspot activity 
and the generation or renewal of polar fields.  A butterfly pattern 
of bipolar sunspot groups is clear from Fig.\ref{butter} as 
their latitudes of emergence shift equatorward with the progression 
of the solar cycle. These bipolar regions constituting leading and 
trailing or following fluxes of opposite polarity are shown in red 
and green in Fig.\ref{butter} for Cycle 21, respectively. 
It is seen from Fig.\ref{butter} that the leading and following polarity 
fluxes in each hemisphere move equatorward and poleward, respectively. 
The leading polarity fluxes are mostly confined to the equatorial 
belt between 0${^{\circ}}$-35${^{\circ}}$. Further, mixed populations 
of the leading polarity fluxes from the opposite hemispheres can be seen 
in the latitude range -10${^{\circ}}$ and 10${^{\circ}}$ that shows the 
cross-equatorial movement of the leading polarity fluxes and their 
eventual cancellation near the equator. 
%
\protect
%

In the meanwhile, the following polarity fluxes in both 
hemispheres drift poleward and eventually reach the poles.  This 
poleward motion of the trailing polarity fluxes can be seen 
from Fig.\ref{butter} as tongues of flux, called polar surges, 
moving poleward in both hemispheres in the latitude band 
45${^{\circ}}$-60${^{\circ}}$. The following polarity fluxes, 
or those moving poleward, eventually reach the poles, cancel 
the previous polar cap fluxes of opposite polarity (red, in 
Cycle 21) and  impart a new polarity at the polar 
cap referred to as polar field reversal. It is, thus, 
clear from Fig.\ref{butter} that the net flux transported 
to the equator or the amount of cross-equatorial flux could 
be correlated with the net flux transported to the poles. Again 
as the net flux transported to the poles is also correlated 
with the net amount of polar cap fields, the net amount of
cross-equatorial flux could be related to polar cap fields.
In this study, we thus investigate the possible correlations 
that could exist between the net flux transport at the equator 
and to the poles, and that between the net flux transport at 
the equator and polar cap fields.
\section{Data and methodology}
We used Carrington synoptic maps available from the magnetic database 
of the National Solar Observatory, Kitt Peak, USA (NSO/KP; 
\url{ftp://nsokp.nso.edu/kpvt/synoptic/mag/}) 
and the Synoptic Optical Long-term Investigation of the Sun facility
(NSO/SOLIS; \url{ftp://solis.nso.edu/synoptic/level3/vsm/merged/carr-rot/}).  
A synoptic Carrington rotation map is generally made 
from several full-disk daily solar magnetograms observed over a 
CR period covering 27.2753 days. These maps, available online as 
standard FITS files contain 180 $\times$ 360 pixels in sine of 
latitude and longitude format.
\protect\begin{figure*}[ht]
\vspace{8.05cm}
\center
\includegraphics{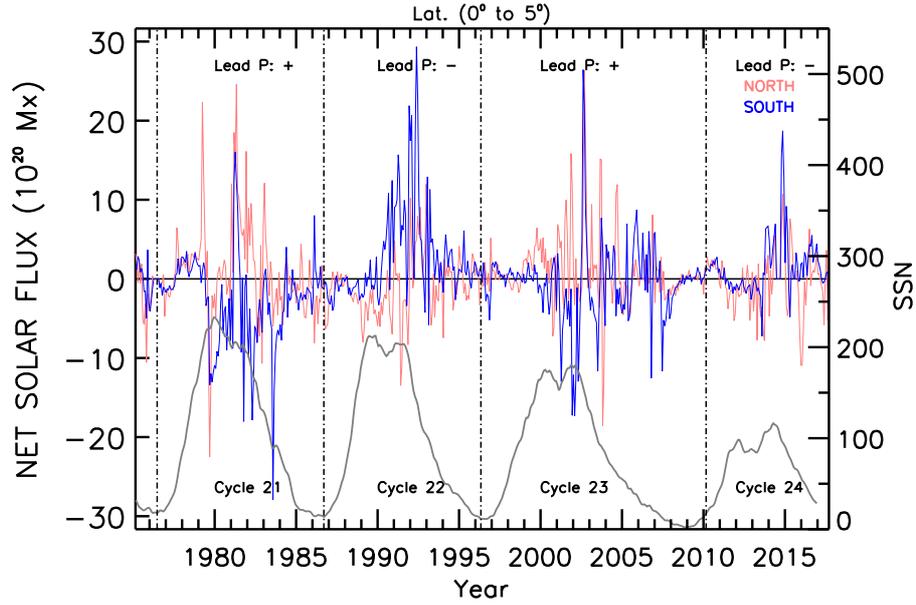}
\caption{The temporal variation of the net solar flux in 
the latitude range, 0${^{\circ}}$-5${^{\circ}}$, in the Northern 
(pink solid line) and Southern (blue solid line) hemispheres 
for Solar Cycles 21--24. Also, the temporal variations in 
SSN are shown by a grey solid line for the 
corresponding periods. The polarity of leading flux 
in the Northern hemisphere is indicated by a positive or 
negative sign at top of the figure. The vertical lines 
are drawn at each solar minimum.}
\label{netf-0} 
\end{figure*}

For each synoptic map, photospheric magnetic fields were first 
estimated  for each of the 180 arrays in sine of latitude by taking 
a longitudinal average of the entire  360 arrays of Carrington 
longitude. Thus, each CR map, previously in the form of an 180 
$\times$ 360 array, is  reduced to an array of 180 $\times$ 1. 
Similarly, such longitudinal strips of 180 $\times$ 1 were obtained 
for all synoptic maps in the period from 1975.14 to 2018.00 and a 
butterfly diagram was generated by placing each strip laterally in 
time.  In short, the magnetic butterfly diagram represents latitudinal 
map of longitudinally averaged solar photospheric fields in time. Care 
was taken to see that data gaps, amounting to $\sim$2.5\% of the 
total data set, were replaced with magnetic field values, estimated 
through a cubic spline interpolation \citep{JBG10}. For achieving a 
better contrast in the butterfly diagram, the field strengths were 
restricted to $\pm30$G, and the two polarities are shown respectively, 
in red (positive) and green (negative) in Fig.\ref{butter}.

For the flux cancellation across the equator we estimate the net 
flux between the latitudes of $\pm$5${^{\circ}}$, while for 
the flux transport to the poles we  estimate the net flux between 
the latitudes of $\pm$45${^{\circ}}$ and $\pm$60${^{\circ}}$. 
The magnetic flux density values for the desired latitude 
ranges were directly estimated from the magnetic butterfly 
diagram shown in Fig.\ref{butter}.  As mentioned earlier, the 
magnetic butterfly diagram comprises of an array of 
180 $\times$573 in sine of latitude and time (in CR). For 
each time (in CR) we have 180 magnetic flux values in 
latitude.  So to obtain the net magnetic flux in time 
for all CRs for a desired latitude range, for e.g. between 
latitude of 0${^{\circ}}$ and 5${^{\circ}}$, the (signed) 
magnetic fluxes are latitudinally averaged in the latitude 
bin of 0${^{\circ}}$ and 5${^{\circ}}$ separately for both 
the Northern and Southern hemispheres.  We have particularly 
stressed in this paper the variation in the net magnetic 
flux at latitudes of 0${^{\circ}}$--5${^{\circ}}$, 
45${^{\circ}}$--50${^{\circ}}$, 50${^{\circ}}$--55${^{\circ}}$, 
and 55${^{\circ}}$--60${^{\circ}}$, respectively. It is 
known \citep{Hat10} and evident from Fig.\ref{butter} that the emergence 
of sunspots is largely confined to the the sunspot 
activity belt at latitudes between $\pm$5${^{\circ}}$ 
and $\pm$35${^{\circ}}$. Therefore, for the estimation of 
flux transport across the equator we have selected the latitudes 
of $\pm$5${^{\circ}}$, where the sunspots are barely seen.  
It is again known \citep{WNS89,Mye12} and evident from Fig.\ref{butter} that 
the flux transport to the poles is mainly confined to the 
latitude range of $\pm$45${^{\circ}}$--$\pm$60${^{\circ}}$ 
where a series of discrete surges or tongues of flux are 
usually observed channelling the magnetic fluxes to the 
poles.  Therefore, for the estimation of the flux transport 
to the poles we have selected the latitude range of 
$\pm$45${^{\circ}}$--$\pm$60${^{\circ}}$.

We estimated the polar cap fields measured above the latitude of 
55${^{\circ}}$, which are actually the line-of-sight magnetic 
fields obtained from the Wilcox Solar Observatory (WSO) 
(\url{http://wso.stanford.edu/Polar.html}). A detail description 
about the WSO polar cap field measurements can be found in 
\cite{JaF18}. While for the monthly averaged total and hemispheric 
smoothed sunspot number (SSN), we used SSN V2.0 observations obtained 
from the Royal Observatory of Belgium, Brussels 
(\url{http://www.sidc.be/silso/datafiles}). The monthly total SSN 
observations are available since 1749, while the monthly 
hemispheric SSN ara available since only 1992. In this study, we 
thus used the monthly total SSN for Cycles 21--24 and monthly 
hemispheric SSN for Cycles 23-24. It is to be noted that a 
recalibrated SSN, known as SSN V2.0, was devised after Jul.
2015 \citep{CLe16,Cli16} and is available at the Royal Observatory 
of Belgium.

\section{Results and Discussions}
  \label{sec:flux-var}
\subsection{Net flux variations at latitude of 0${^{\circ}}$- $\pm$5${^{\circ}}$}
The net magnetic flux in the latitude band of 0${^{\circ}}$-5${^{\circ}}$ 
is shown in Figure \ref{netf-0}. The pink and blue solid lines respectively 
represent temporal variation of magnetic measurements in the Northern and 
Southern hemispheres. The polarity of leading sunspot in the Northern 
hemisphere in each
cycle is shown by a positive (+) or negative (-) sign at top in pink 
(parameters in the North are represented in pink and the South 
in blue). The net magnetic flux at the beginning of a cycle is 
generally less.   This can be simply attributed to lower flux 
transport from the higher solar latitudes because at the start 
of the solar cycle sunspots have just started emerging at latitudes 
of $\pm$35${^{\circ}}$ in both hemispheres.  The net flux then rises 
up gradually as the flux starts getting transported from the higher 
latitudes, and at the solar cycle maximum, the net flux is at the 
peak. Afterward, in the declining phase of the cycle, the 
net flux dips down because of the reduction in the number of 
sunspots, and the rise in flux cancellation at the equator.  It 
is clearly seen from Figure \ref{netf-0} that the polarity of the 
net magnetic flux in both the hemispheres is same as that of the 
polarity of the leading sunspots in their respective 
hemispheres. 
\protect\begin{figure*}[ht]
\vspace{8.05cm}
\center
\includegraphics{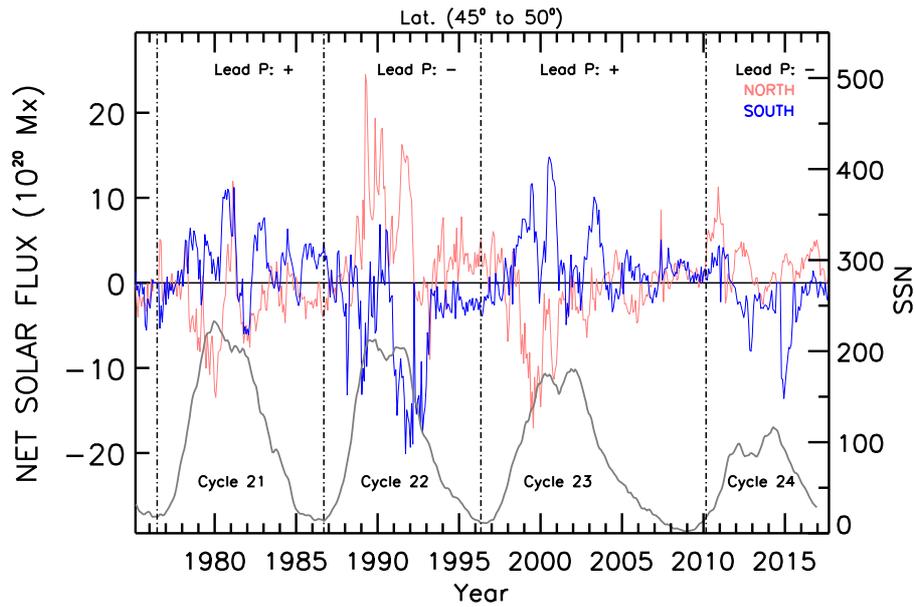}
\caption{The temporal variation of the net solar flux in 
latitude range, 45${^{\circ}}$-50${^{\circ}}$, for the Northern 
and Southern hemispheres marked in pink and blue, respectively, 
shown for Solar Cycles 21--24.}
\label{netf-45} 
\end{figure*}

From the 
variation of the net hemispheric flux in the latitude band of 
0${^{\circ}}$- $\pm$5${^{\circ}}$ the presence of the wrong 
orientation fluxes are also easily observable. As described earlier, 
we see that the bipolar sunspot regions generally follow the 
Hale's polarity law \citep{HaE19} so that the leading and following spots have 
a polarity opposite to that of the corresponding polarity of the 
leading and following spots in the other hemisphere. However, it is
sometimes found that bipolar regions have an orientation 
opposite to the normally expected orientation of bipolar regions.  
Such fluxes are referred to as wrong orientation fluxes \citep{WNS89,JiC15}. 
For example, in Cycle 21, we can clearly see sudden changes in 
the polarity of the net magnetic flux in the Northern hemisphere 
from positive to negative during the period around 1980. We also 
find a similar behaviour in the Southern hemisphere around the period 
1982. Similarly, 
wrong orientation fluxes are also observed in Cycle 23 where the 
polarity of the net flux in the Southern hemisphere changes from 
positive to negative around 2004. Such changes in the polarity of 
net fluxes indicate the appearance of low latitude bipolar 
regions having the wrong orientation of fluxes.

A close inspection of the net flux variation in Cycle 24 shows 
a peculiarity in the Southern hemisphere. As expected, the 
polarity of the net flux in the beginning of the cycle was positive 
until 2012 while it was unexpectedly negative from 
2012--2014, and only changed back to normal polarity 
when it approached solar cycle maximum. However, such behaviour 
of the net flux has not been noticed for other Solar Cycles 21--23 
from the Southern hemisphere. Also, a comparison of the net magnetic 
flux variation in the Northern and Southern hemispheres in the last 
four Solar Cycles 21--24 shows a clear hemispheric asymmetry.  
However, it is important to note that this hemispheric asymmetry 
is much more significant in Cycle 22 as compared to other solar 
cycles wherein, the flux from the Southern hemisphere is 
seen significantly higher than that of the Northern hemisphere 
during the solar maximum of the cycle. 
A quantification of the hemispheric magnetic flux asymmetry is 
further discussed in Section \ref{sec:correlation}.
\protect\begin{figure*}[ht]
\vspace{8.05cm}
\center
\includegraphics{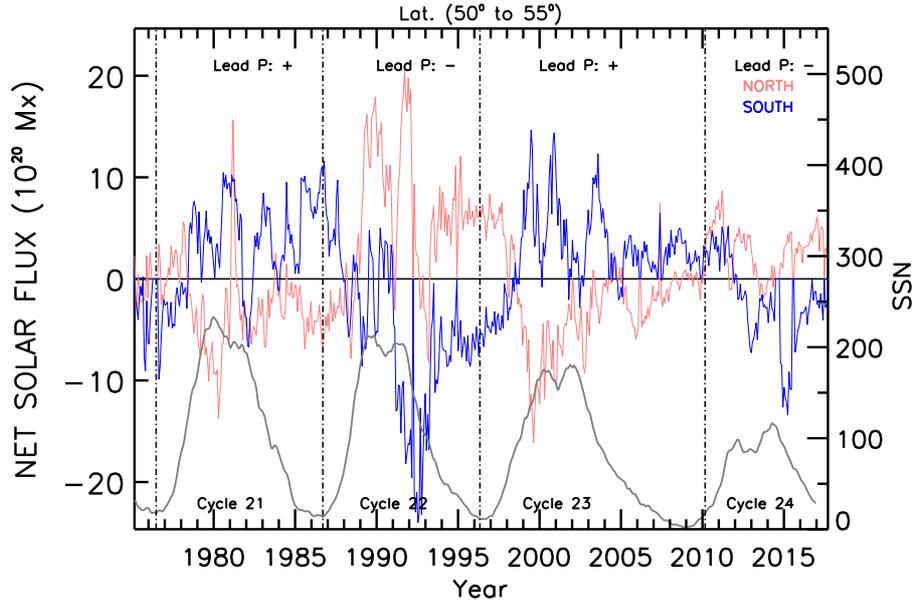}
\caption{The temporal variation of the net solar flux in 
latitude range, 50${^{\circ}}$-55${^{\circ}}$, for the Northern 
and Southern hemispheres marked in pink and blue, respectively, 
shown for Solar Cycles 21--24.}
\label{netf-50} 
\end{figure*}
%
%
\subsection{Net flux variations at latitude of 
$\pm$45${^{\circ}}$- $\pm$50${^{\circ}}$}\label{sec:netf-45}
The net flux in the latitude band of 
$\pm$45${^{\circ}}$--$\pm$50${^{\circ}}$ is shown in Figure 
\ref{netf-45} for both the Northern and Southern hemispheres. 
Unlike the latitude band of 0${^{\circ}}$-5${^{\circ}}$, the 
polarity of the net flux in each hemisphere is opposite to 
that of the polarity of leading sunspots.  However, as 
mentioned for the latitude band of 
0${^{\circ}}$--$\pm$5${^{\circ}}$, for this latitude band 
we also found changes in the polarity of the net flux.  In 
particular it is seen that, such changes are comparatively 
more frequent in Cycles 21 and 23 compared to that of 
Cycles 22 and 24.  In Cycle 21 a sudden large change in the 
polarity of the net flux is found in the North around 1980, 
while a similar change in the polarity of the net flux is 
observed at around 1982 in the South.  These changes in the 
polarity of the net flux corresponds to the same period of 
time when changes in the polarity of the net flux have been 
observed in the latitude band of 0${^{\circ}}$- $\pm$5${^{\circ}}$.  
Thus, this further confirms the low latitude emergence of 
the bipolar magnetic regions with the wrong orientation.  
Similarly, we have also noticed, in Cycle 22, random changes 
in polarity of the flux in the South around the solar cycle 
maximum while the same has been noticed frequently in Cycle 
23 both in the Northern and Southern hemispheres after the solar 
cycle maximum.  In addition, the strength of the net hemispheric 
flux in Cycle 23 was comparatively less after around 2004,
\protect\begin{figure*}[ht]
\vspace{8.05cm}
\center
\includegraphics{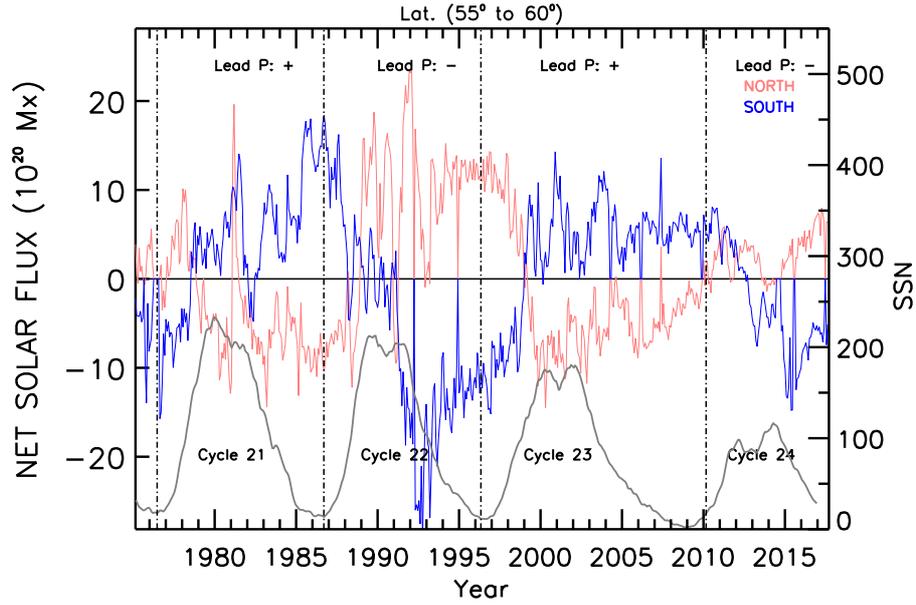}
\caption{The temporal variation of the net solar flux 
in the latitude range, 55${^{\circ}}$-60${^{\circ}}$, for 
both solar hemisphere, with the Northern and Southern 
hemispheres marked in pink and blue, respectively, shown 
for Solar Cycles 21--24.}
\label{netf-55} 
\end{figure*}
%
and was subsequently reduced during the prolonged minimum of 
Cycle 23 indicating that the flux transport to the polar cap region 
was low. This may explain the weakness of polar field strength 
during the minimum of Cycle 23.  Unlike the latitude band of 
0${^{\circ}}$-$\pm$5${^{\circ}}$ the net flux in Cycle 24 
shows a normal or expected behaviour though its strength is 
comparatively weaker.

Also, from Fig.\ref{netf-45}, the hemispheric asymmetry in the 
net magnetic flux variation in each solar cycle is clearly evident 
and like the latitude band of 0${^{\circ}}$--$\pm$5${^{\circ}}$, 
the hemispheric asymmetry is more significant in Cycle 22. In addition, 
the same in Cycle 24 also shows significant difference during the 
maximum of Cycle 24 with the flux for the Southern hemisphere higher than that 
of the Northern hemisphere.
%
\begin{table*}[ht]
\begin{center}
\begin{tabular}{llllll}
\hline
&Latitude &Solar Cycle &${\lvert \phi \rvert}_{N}$ ($\times$10$^{22}$Mx) &${\lvert \phi \rvert}_{S}$ ($\times$10$^{22}$Mx) &${\Gamma_{\Phi}}$\\
\hline
&0${^{\circ}}$--5${^{\circ}}$   &21  &1.29  &3.61  &0.47\\
&                               &22  &1.45  &3.36  &0.40\\
&                               &23  &1.93  &1.21  &0.23\\
&                               &24  &0.49  &1.33  &0.49\\
\hline
&45${^{\circ}}$--50${^{\circ}}$ &21  &3.00  &3.07  &0.01\\
&                               &22  &5.53  &5.69  &0.01\\
&                               &23  &3.03  &3.97  &0.13\\
&                               &24  &2.30  &1.80  &0.12\\
\hline
&50${^{\circ}}$--55${^{\circ}}$ &21  &3.68  &4.29  &0.07\\
&                               &22  &6.84  &6.76  &0.01\\
&                               &23  &3.70  &4.54  &0.11\\
&                               &24  &2.80  &2.25  &0.10\\
\hline
&55${^{\circ}}$--60${^{\circ}}$ &21  &5.05  &5.52  &0.04\\
&                               &22  &8.57  &8.06  &0.03\\
&                               &23  &4.34  &5.02  &0.07\\
&                               &24  &2.71  &2.60  &0.02\\
\hline
\end{tabular}
\end{center}
\caption{Lists the estimated net flux values for Solar Cycles 
21--24 in the Northern (${\lvert \phi \rvert}_{N}$) and 
Southern (${\lvert \phi \rvert}_{S}$) hemispheres and the 
hemispheric magnetic flux asymmetry (${\Gamma_{\Phi}}$) in the 
latitude bands 0${^{\circ}}$- $\pm$5${^{\circ}}$, 
$\pm$45${^{\circ}}$- $\pm$50${^{\circ}}$, 
$\pm$50${^{\circ}}$- $\pm$55${^{\circ}}$, 
and $\pm$55${^{\circ}}$- $\pm$60${^{\circ}}$, respectively.}
\label{tab-1}
\end{table*}
\subsection{Net flux variations at latitude of 
$\pm$50${^{\circ}}$- $\pm$55${^{\circ}}$}\label{sec:netf-50}
The temporal variation in the net flux in latitude bin 
50${^{\circ}}$-55${^{\circ}}$ is shown in Figure 
\ref{netf-50} both in the Northern and the Southern hemispheres. 
The variations seen show that they are almost similar 
in behaviour to the net flux variations seen in the 
latitude bin, 45${^{\circ}}$-50${^{\circ}}$.   It is 
seen that the polarity of the net flux also behaves 
in the same way. The difference that distinguishes 
it from the previous latitude bin is that the strength 
of the net flux in every cycle is comparatively higher.  
Also, the 
abrupt changes in the polarity of the net flux that are 
seen during Solar Cycle 22 are for a longer duration than 
earlier and the strength of the net flux in the South is stronger. 
However, unlike the frequent changes of the polarity of the 
net flux (as seen in Cycles 21 and 23 in the previous latitude 
band of 45${^{\circ}}$ - 50${^{\circ}}$) we find such changes 
are less frequent in this latitude band. This indicates that 
both polarity fluxes are transported up to the latitude of 
$\pm$50${^{\circ}}$.  
%
\begin{table*}[ht]
\begin{center}
\begin{tabular}{llllll}
\hline
&\multicolumn {2}{c}{Latitude} &Hemisphere  &Pearson's Correlation &Spearman's Correlation \\
\hline
&0${^{\circ}}$--5${^{\circ}}$  &45${^{\circ}}$--50${^{\circ}}$ &North  &-0.26(P=99\%)	  &0.50(P=66\%)  \\
&                              &                               &South  &0.08 (P=99\%)	  &-0.50(P=66\%)  \\
&                              &                               &North \& South  &0.10 (P=99\%)	  &0.20(P=70\%)  \\
\hline
&0${^{\circ}}$--5${^{\circ}}$  &50${^{\circ}}$--55${^{\circ}}$ &North  &-0.27 (P=99\%)	&0.5(P=66\%) \\
&                              &                               &South  &0.33 (P=99\%)	  &-0.5(P=66\%) \\
&                              &                               &North \& South  &0.21 (P=99\%)	  &0.08(P=87\%)  \\
\hline
&0${^{\circ}}$--5${^{\circ}}$  &55${^{\circ}}$--60${^{\circ}}$ &North  &-0.42 (P=99\%)	&-0.50(P=66\%)  \\
&                              &                               &South  &0.55 (P=99\%)	  &0.5(P=66\%)  \\
&                              &                               &North \& South  &0.22 (P=99\%)	  &0.31(P=54\%)  \\
&0${^{\circ}}$--5${^{\circ}}$  &55${^{\circ}}$--90${^{\circ}}$ &North  &-0.99 (P=99\%)	&-1.00(P=66\%)  \\
&                              &                               &South  &0.87 (P=99\%)	  &1.00(P=66\%)  \\
&                              &                               &North \& South  &0.51 (P=99\%)	  &0.31(P=54\%)  \\
\hline
\end{tabular}
\end{center}
\caption{Lists the Pearson's and Spearman's rank correlations between the net flux values in the latitude bands of 
0${^{\circ}}$-5${^{\circ}}$ and 45${^{\circ}}$-50${^{\circ}}$, 0${^{\circ}}$-5${^{\circ}}$ and 50${^{\circ}}$-55${^{\circ}}$, 
0${^{\circ}}$-5${^{\circ}}$ and 55${^{\circ}}$-60${^{\circ}}$, 0${^{\circ}}$-5${^{\circ}}$and 55${^{\circ}}$-90${^{\circ}}$ 
for the Northern, Southern, and both hemispheres, estimated for Solar Cycles 21--23.}
\label{tab-2}
\end{table*}
%

The hemispheric asymmetry in the net flux variation in each 
solar cycle for this latitude band shows a similar behaviour 
like that in the latitude band of 45${^{\circ}}$--50${^{\circ}}$.  
\subsection{Net flux variations at latitude of 
$\pm$55${^{\circ}}$- $\pm$60${^{\circ}}$}\label{sec:netf-55}
Figures \ref{netf-55} represents the net flux for both the 
Northern and Southern hemispheres in the latitude band of 
$\pm$55${^{\circ}}$-$\pm$60${^{\circ}}$.  The net flux in 
this latitude bin, as depicted in Figure \ref{netf-55}, 
is actually more like the polar cap flux, which shows a 
reversal in polarity of the flux in each cycle at around 
solar cycle maximum, unlike the net flux in the previous 
latitude bands.  At the initial phase of solar cycle, the 
net hemispheric flux has the same polarity to that of the 
leading sunspot flux.  As the solar cycle progresses the 
net hemispheric flux strength decreases and at around the 
solar maximum the polarity of the net hemispheric flux 
reverses, and is the same as that of the polarity of the 
following sunspot flux. 
 
Actually, as pointed out previously, the surplus amount 
of trailing polarity fluxes transported from the low 
latitudes, on reaching in this regime, cancel the old 
polar cap flux to produce the new polar cap flux having 
the opposite polarity.  A comparison of the strength of 
the net flux in this latitude band for Cycles 21\,--\,24 
shows that the net flux was weaker during the 
prolonged minimum of Cycle 23 as compared to the earlier 
minima of Solar Cycles 20--22.  The net flux in the 
rising phase of the Cycle 24 as well as in the declining 
phase of the Cycle 24 were also found weaker in strength 
as compared to the earlier cycles. Also, the hemispheric 
asymmetry in the net flux variation for each solar cycle, 
like in the previous latitude bands of 
45${^{\circ}}$--50${^{\circ}}$ and 50${^{\circ}}$-55${^{\circ}}$, 
is clearly seen during the maxima of Cycles 22 and 24.
%
%
\protect\begin{figure*}[ht]
\vspace{13.65cm}
\center
\includegraphics{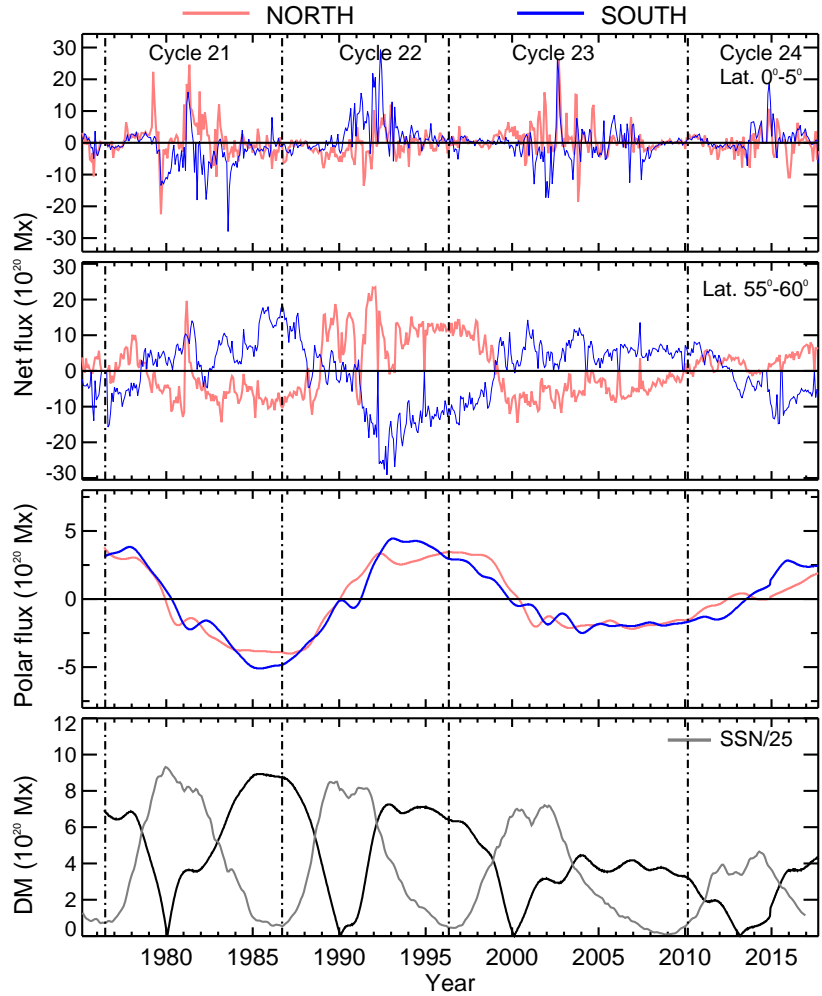}
\caption{(first and second panels) the temporal variation in 
the net solar flux in latitude range, 0${^{\circ}}$-5${^{\circ}}$ 
(first panel) and 55${^{\circ}}$-60${^{\circ}}$ (second panel) 
for the northern and southern hemispheres marked in pink 
and blue respectively for Solar Cycles 21-23. (third panel) 
the variation in the north polar cap flux (pink) and 
negative of south polar cap flux (blue) estimated from the 
WSO polar field measurements for the last four solar cycles. 
(fourth panel) the variation of axial dipole moment estimated 
from WSO polar fields (black) overplotted by the monthly 
averaged SSN (grey) divided by 25. The vertical lines demarcate 
the solar minima differentiating the solar cycles.}
\label{netflux-0n55}
\end{figure*}
%
\subsection{Correlation of the net solar fluxes} 
\label{sec:correlation}
From the investigation of variations in the net flux at 
different latitude bands, as described above, it is 
apparent that the flux cancelled across the equator 
and the flux transported to the poles may have a 
correlation.  For finding 
the correlation between the net amount of flux in the latitude belt 
of 0${^{\circ}}$--$\pm$5${^{\circ}}$ with that of the corresponding 
net amount of flux in the latitude band $\pm$45${^{\circ}}$--$\pm$50${^{\circ}}$, 
$\pm$50${^{\circ}}$--$\pm$55${^{\circ}}$, and $\pm$55${^{\circ}}$--$\pm$60${^{\circ}}$, 
respectively, the net amount of flux in both the North and the South is 
estimated for each solar cycle from Cycles 21\,--\,24. The 
estimated fluxes in both the North and South for each solar 
cycle are listed in Table \ref{tab-1} for these latitude bands. 
We also estimated the magnetic flux asymmetry, ${\Gamma_{\Phi}}$:

\begin{equation}
  \Gamma{_{\phi}}= \frac{{\lvert \phi \rvert}_{N} - {\lvert \phi \rvert}_{S}}{{\lvert \phi \rvert}_{N} + {\lvert \phi \rvert}_{S}}
 \label{eqphi}
 \end{equation}
where ${\lvert \phi \rvert}_{N}$ and ${\lvert \phi \rvert}_{S}$ are 
absolute values of the net flux for the Northern and Southern 
hemispheres. The estimated values of ${\Gamma_{\Phi}}$ in each solar 
cycle are also listed in Table \ref{tab-1} for these latitude bands.

An inspection of the net flux values as shown in 
Table \ref{tab-1} reveals that the net amount of 
flux in the latitude band 0${^{\circ}}$--$\pm$5${^{\circ}}$ 
shows a progressive increase in its strength from Cycle 
21 to 23 in the North while for the South it shows a 
reduction in its strength. However such behaviour has 
not been observed for all the other latitude bands, 
although it is noticed that the net flux shows a 
progressive decrease in its strength from Cycle 22 to 
24 for both the North and South with the flux is higher 
in Cycle 22 in comparison to Cycles 21, 23 and 24. This 
suggests a continuous weaker polar flux since Cycle 22.
It is to be reminded here that in our earlier studies 
\citep{JBG10,JaB15a,IJB19} we also showed a similar steady 
decrease in unsigned polar field strength since Cycle 
22.
While a comparison of the net flux values for the North and 
South in the latitude band 0${^{\circ}}$--$\pm$5${^{\circ}}$ 
shows a weaker net magnetic flux for the North in each solar 
cycle except for Cycle 23 wherein the net magnetic flux in 
the North is stronger than the South. Also, a lower value 
(0.23) of ${\Gamma_{\Phi}}$ has been found for Cycle 23 as 
compared to other Solar Cycles (0.40-0.49). On the other 
hand, the comparison of the net flux values for the North 
and South in the latitude band 45${^{\circ}}$--50${^{\circ}}$ 
shows a weaker net magnetic flux for the North in each 
solar cycle except for Cycle 24. In the latitude bands 
50${^{\circ}}$--55${^{\circ}}$ and 55${^{\circ}}$--60${^{\circ}}$, 
the net flux values for the North and South are alternatively 
stronger or weaker from Cycles 21--24. While the inspection of 
${\Gamma_{\Phi}}$ values for these latitude bands shows no 
specific trend. However, the value of ${\Gamma_{\Phi}}$ 
has been found to be comparatively higher for Cycle 23 as 
compared to other Solar Cycles.

We estimated the Pearson's correlation coefficient 
and the Spearman's rank correlation coefficient by 
comparing the net amount of flux in the Northern 
hemisphere in the latitude belt of 
0${^{\circ}}$--$\pm$5${^{\circ}}$ with that of 
of the net flux in the latitude belt of 
$\pm$45${^{\circ}}$--$\pm$50${^{\circ}}$, 
$\pm$50${^{\circ}}$--$\pm$55${^{\circ}}$, 
and $\pm$55${^{\circ}}$--$\pm$60${^{\circ}}$, 
respectively.  Similarly, we repeated the 
same for the Southern hemisphere too.  All these estimated 
coefficient values are listed in Table \ref{tab-2} with their 
statistical significance. It is evident form Table \ref{tab-2} 
that a moderately good negative correlation (Pearson's correlation 
coefficient, r = -0.42, P=99\% confidence level) is observed 
between the net flux in the latitude band of 
0${^{\circ}}$--5${^{\circ}}$ and 55${^{\circ}}$--60${^{\circ}}$ 
for the North while a moderately good positive correlation (Pearson's 
correlation coefficient, r = 0.55, P=99\%) is observed for 
the South. It thus indicates that the flux cancellation at 
the equator and the flux transport to the poles have a good 
correlation.
When we consider the net flux values of the North and the 
South together to find any correlation between the latitude 
band of 0${^{\circ}}$--5${^{\circ}}$ and 
55${^{\circ}}$--60${^{\circ}}$, we found low values of 
Pearson's correlation coefficient (r=0.22, P=99\%) and 
Spearman's rank correlation coefficient ($\rho$=0.31, P=54\%). 
Thus, it is clear that the correlation for the net flux 
only exists when the North and South are taken separately, 
but not for the net flux for the North and South considered 
together. 

Taking advantage of the new correlation between the net 
flux cancelled at the equator and the net flux transported 
to the poles, we tried to see if any correlation exists 
between the net flux cancelled at the equator and the polar 
fields at solar minimum. This has been discussed in the 
following section.
\subsection{Correlation of the net flux at the equator and 
the polar cap fields} 
  \label{sec:flux-cantr}
For an easier comparison of the net flux in the latitude band of 
0${^{\circ}}$--5${^{\circ}}$ and 55${^{\circ}}$--60${^{\circ}}$ during 
the past four Solar Cycles 21--24, we plot the respective net flux, 
obtained from NSOKP magnetic field measurements, in the first and 
second panels of Figure \ref{netflux-0n55} for both the Northern (pink) 
and Southern (blue) hemispheres, respectively. The variation in the polar 
cap flux, obtained from WSO measurements, is shown in the third panel of 
%
\begin{table*}[ht]
\begin{center}
\begin{tabular}{llllll}
\hline 
&Solar Cycle  &Year  &CR  &PMF${_{min}^{n}}$ (N)  &PMF${_{min}^{n}}$ (S) \\
\hline
&21  &1986      &1771-1783  &-8.5	&10.7 \\
&22  &1996      &1905-1917  &7.2	&-6.9 \\
&23  &2008-2009 &2072-2084  &-3.6	&4.0  \\
\hline
\end{tabular}
\end{center}
\caption{Lists the estimated values of polar cap fields (PMF${_{min}^{n}}$) in 
the Northern and Southern hemispheres during the solar minima of Cycles 21--23. 
The values of polar fields are in unit of Gauss.}
\label{tab-3}
\end{table*}
%
Fig.\ref{netflux-0n55}. The polar cap fields for the Southern 
hemisphere (blue) have been inverted in order to compare them with 
the polar cap fields for the Northern hemisphere (pink).  While 
the axial dipole moment, estimated from the absolute difference 
between the Northern and Southern WSO polar cap fields, is shown 
in the fourth panel. Over-plotted in the fourth panel 
is also the monthly averaged SSN in grey, which is divided by 25 
to fit into the Y-axis.  The dotted vertical lines in each panel 
of Fig.\ref{netflux-0n55} demarcate the solar minima of Cycles 20--23. 
As mentioned earlier, it is clear from the first and second panels 
of Fig.\ref{netflux-0n55} that the net flux at the equator and the 
poles in Cycle 22 is comparatively more than that in cycles 21 
and 23. While the net flux in Cycle 24 both at the equator and 
the poles is lower than that of Cycle 23. The same is found for the 
polar cap field strength during solar minimum of each cycle (third panel of 
Fig.\ref{netflux-0n55}) with the field strength in Cycle 24 showing 
weaker values than that in Cycle 23. Thus, there may be a 
correlation between the net flux at the equator during each 
cycle (EMF${^{n}}$) and the polar cap fields at 
cycle minimum (PMF${_{min}^{n}}$).

\begin{figure*}[ht]
\vspace{6.65cm}
\center
\includegraphics{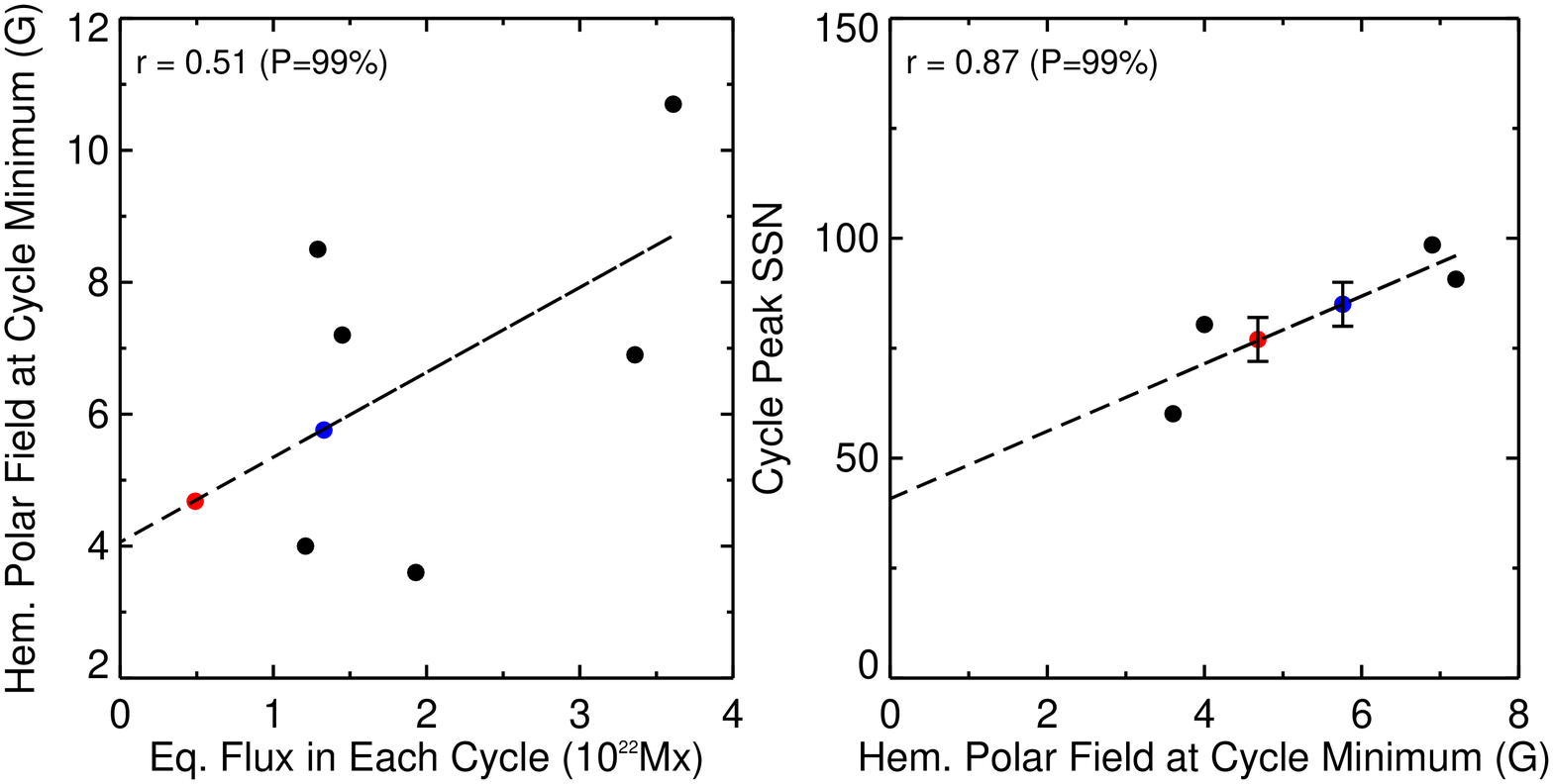}
\caption{(Left) correlation between net equatorial flux of each 
cycle with polar field at solar minimum shown with r=0.51 (P=99\%). 
The filled black dots are data points for the Northern and Southern 
hemispheres covering Solar Cycles 21--23. The dashed line corresponds 
to a best fit obtained using the least square residual method. 
The filled red and blue dots are, respectively, values for the 
Northern and Southern hemisphere for Cycle 24. (Right) correlation 
between polar field at solar minimum of a preceding cycle and 
cycle peak of the following cycle shown with r=0.87 (P=99\%). The 
filled black dots are data points for the Northern and Southern 
hemispheres for Cycles 22/23--23/24, while the filled red and blue 
dots represent the data point for the Northern and Southern 
hemispheres, respectively, for Cycle 24/25. The dashed line 
is a best fit to all the data points.}
\label{corr}
\end{figure*}

In order to find the correlation, we estimated WSO polar cap field strength 
during each solar cycle minimum, from Cycles 21--23, by averaging the values 
of polar cap fields in each hemisphere for an interval of one year around 
each solar cycle minimum. The one year intervals \citep{WRS09} around 
the minima of Solar Cycles 21, 22, and 23 were chosen corresponding 
to CR periods of CR1771--1783, CR1905--1917, and 
CR2072--2084, respectively. The estimated values of hemispheric polar cap 
fields for Cycles 21--23 are listed in Table \ref{tab-3}.  We have
then compared these values with the already estimated values of 
the net flux in the latitude band 0${^{\circ}}$--5${^{\circ}}$, 
those representing the flux cancelled at the equator, and found a good 
correlation with a Pearson's correlation coefficient of r = 0.51 and P = 99\% 
confidence level.  Shown in the left panel of Figure \ref{corr} is the 
correlation of EMF${^{n}}$ versus PMF${_{min}^{n}}$. The filled black dots in 
Fig.\ref{corr} (left) represent data points for both the Northern and Southern 
hemispheres for Cycles 21--23. The solid black line in Fig.\ref{corr} is a 
best fit to all the data points obtained using a least square residual method. 
The linear correlation of EMF${^{n}}$ and PMF${_{min}^{n}}$ can be represented 
by the following equation,
 \begin{equation}
 PMF{_{min}^{n}} = 4.06\pm(0.2) + (1.28\pm0.05) \times EMF{^{n}}
 \label{eq1}
 \end{equation}
%
\protect
Using the values of EMF${^{n}}$ of 0.49 $\times$ 10${^{22}}$ Mx and 
1.33 $\times$ 10${^{22}}$ Mx for the Northern and Southern hemispheres, 
respectively, for Cycle 24 (see Table \ref{tab-1}) in Equation \ref{eq1}, 
we estimated the values of PMF${_{min}^{n}}$ of 4.68 G and 5.76 G for the 
Northern and Southern hemispheres, respectively, for Cycle 24. The value 
for the North is indicated by a solid red circle in Fig.\ref{corr} (left), 
while the value for the South is indicated by a solid blue circle.

An interesting feature in the correlation, when it is 
estimated separately for the North and South, is an 
improvement in the Pearson's correlation coefficient.  
So for the North we found a Pearson's correlation 
coefficient of r = -0.99 and P = 99\%, while for the 
South we found a Pearson's correlation coefficient 
of r = 0.87 and P = 99\%.
\subsection{Amplitude of solar cycle 25}\label{sec:ampl25}
Using polar flux measurements for Solar Cycles 14--23, \cite{MBD13} 
showed that the solar cycle predictions made separately for each hemisphere 
provide a better estimate of SSN${_{max}}$ of the next cycle. Taking into 
this account, we used the estimated values of hemispheric
PMF${_{min}^{n}}$ at Cycle 24 minimum to predict the amplitude 
of Solar Cycle 25 for the Northern and Southern hemispheres. We thus 
derived a correlation equation between 
hemispheric PMF${_{min}^{n}}$ and SSN${_{max}^{n+1}}$ using hemispheric 
measurements of PMF${_{min}}$ and SSN${_{max}}$ obtained for Solar 
Cycles 22 and 23, as shown in the right panel of Fig.\ref{corr}. The 
filled solid dots in Fig.\ref{corr} (right) represent the data points 
for Cycles 22-23.  Although there are only four points, we found a 
very good statistically significant correlation 
between PMF${_{min}^{n}}$ and SSN${_{max}^{n+1}}$ with a Pearson correlation 
coefficient of $r = 0.87$ at a significance level of $99\%$. Also, it is to 
be noted that the polar field strength at the minimum of a preceding 
cycle has been successfully used as a precursor for predicting the amplitude 
of the next cycle \citep{ScS78,SPe93,Sch05} wherein the authors used either 
two or three data points only. The solid black 
line in Fig.\ref{corr} (right) is a best fit to all the data points 
between PMF${_{min}^{n}}$ and SSN${_{max}^{n+1}}$. 
The linear correlation between them can be represented by the following equation,

 \begin{equation}
 SSN{_{max}^{n+1}} = 40.78\pm(17) + (7.67\pm3.0) \times PMF{_{min}^{n}}
 \label{eq2}
 \end{equation}

Using the value of PMF${_{min}^{n}}$ of 4.68 G for the Northern 
hemisphere for Cycle 24 (see Section \ref{sec:flux-cantr}) in 
equation \ref{eq2}, we predict a SSN${_{max}}$ of 77$\pm$5 
for Solar Cycle 25. This value is indicated by a solid red circle with a 
1 sigma error-bar in Fig.\ref{corr} (right). Similarly, using the 
value of PMF${_{min}^{n}}$ of 5.76 G for the Southern 
hemisphere for Cycle 24 in equation \ref{eq2}, we predict a 
SSN${_{max}}$ of 85$\pm$5 for Solar Cycle 25. This value is indicated by 
a solid blue circle with 1 sigma error-bar in 
Fig.\ref{corr} (right).
\section{Summary and Conclusions}\label{sec:con}
In this paper, we have used photospheric magnetic field measurements from 
NSOKP covering a period from Solar Cycles 21\,--\,24 to study and compare 
solar magnetic flux  cancelled at the equator and transported to the poles. 
For the flux cancelled at the equator, we have estimated the net flux in the 
latitude belt $0^{\circ}$--$5^{\circ}$, while for the flux transported 
to the poles we have estimated the net flux in the latitude 
belt $45^{\circ}$--$50^{\circ}$, 
$50^{\circ}$--$55^{\circ}$, and $55^{\circ}$--$60^{\circ}$. 
We have estimated the net flux for each latitude belt for both the Northern 
and Southern hemispheres for each solar cycle from Cycles 21--24.  The 
net flux variations in each latitude belt show different behaviour 
in each solar cycle, although an hemispheric asymmetry in the net 
flux strength is seen for each solar cycle in all the latitude belts. 
This asymmetry is more pronounced during the maximum of Cycle 22 as 
compared to the maxima of other cycles. While the value of hemispheric 
magnetic flux asymmetry for the latitude belt $0^{\circ}$--$5^{\circ}$ 
shows a lower value for Cycle 23 as compared to other solar cycles, the 
value of the same for all other latitude belts shows a higher value for 
Cycle 23 as compared to other solar cycles. Compared to earlier cycles 
both the flux cancelled at the equator and transported to the poles, 
in Cycle 24, show weaker hemispheric flux strength.

It is usually seen that the overall polarity of the net flux in 
each hemisphere is same as that of the leading sunspot polarity 
for the low latitude belt, while for the high latitude belt it is 
same as that of the following sunspot polarity.  However, sometimes 
around the maximum of solar cycle, we have observed the change 
of polarity in the net flux transport at the equator, called as the 
wrong orientation fluxes, due to the emerging of low latitude rogue 
active regions. The same is also observed in the net flux estimated 
for the high latitude belts. This change of polarity due to the 
wrong orientation fluxes is different from the change in 
polarity of solar polar fields that usually always occurs during 
the maximum of each solar cycle. Such wrong orientation fluxes have 
been mostly seen in Cycles 21 and 23. It is argued \citep{JiC15} 
that the wrong orientation fluxes seen 
during Cycle 23 have contributed to the weakness of Solar Cycle 24. 
Recently, \cite{JaF18} reported the unusual polar field reversal 
pattern in Cycle 24 that showed a prolonged near-zero polar magnetic 
field condition in the Northern hemisphere after the start of the polar reversal. 
The multiple wrong orientation fluxes seen in Cycle 24, 
from this study, corresponding to the above period can explain the 
prolonged near-zero magnetic field condition in the Northern 
hemisphere.

Further, we have found a good correlation between the net flux 
in the latitude belt of $0^{\circ}$--$5^{\circ}$ and $55^{\circ}$--$60^{\circ}$
for the Northern as well as for the Southern hemispheres, and a 
much better correlation when we considered the net flux for 
both the Northern and Southern hemispheres together. Next, we have 
also compared the flux cancelled at the equator during each cycle to the polar cap field 
strength at the cycle minimum. We have found a good correlation between them when 
considered separately for the Northern and Southern hemispheres and together. We have 
used this correlation to estimate the hemispheric polar field strength at the Solar 
Cycle 24 minimum. The values, in turn, are used to estimate SSN${_{max}}$ of 77$\pm$5 
and 85$\pm$5 for the Northern and Southern hemispheres, respectively, for the upcoming 
Solar Cycle 25. This is actually in agreement with the prediction made by \cite{GoM18} 
wherein, the authors predicted SSN${_{max}}$ of 59 and 89 for the Northern and Southern 
hemispheres for Solar Cycle 25, respectively. The predicted values, from this study, 
for the amplitude of Solar Cycle 25 thus indicates that we would witness another 
mini-solar maximum, like that in Cycle 24, in the upcoming Solar Cycle 25. 
\cite{JaF18} reported (see the upper 
panel of Fig.7 in the paper), that the strength of unsigned solar polar magnetic fields 
in the Southern hemisphere is relatively stronger than the Northern hemisphere in Solar 
Cycle 24. Thus, the amplitude of Solar Cycle 25 for the Southern 
hemisphere could be stronger than that for the Northern hemisphere 
that is exactly what our study reports here. In addition, it is shown 
in Fig.\ref{netflux-0n55} (third panel) of this study that an 
hemispheric asymmetry exists for the signed solar polar fields in 
Cycle 24 with the Southern polar fields reversing earlier than the 
Northern polar fields. This could lead to a delay in the timing of the 
sunspot minimum of Cycle 24 for the Northern hemisphere. 
As a result, there could be an hemispheric asymmetry in the timing 
of the sunspot maximum for the Northern and Southern hemispheres 
in the upcoming Solar Cycle 25. 

In the present space age, the study of such evolution of 
solar photospheric magnetic flux and solar cycle predictions are 
important to plan the mission years of space and 
planetary exploration programs. Continued investigations of 
solar magnetic flux should thus be carried out.
\acknowledgement
The authors would thank the free use data policy of the National Solar
Observatory (NSO/KP and NSO/SOLIS) and the Wilcox Solar Observatory (WSO) 
for the synoptic magnetogram data and World Data Center SILSO at Royal Observatory, 
Belgium, Brussels for the sunspot data. The SOLIS data 
obtained by the \emph{NSO Integrated Synoptic Program} (NISP), managed 
by the National Solar Observatory, which is operated by the Association of 
Universities for Research in Astronomy (AURA), Inc. under a cooperative agreement 
with the National Science Foundation. Data storage supported by the University of 
Colorado Boulder ``PetaLibrary.'' 
SKB acknowledges Dr. Dibyendu Nandi from Indian Institute of Scientific and Educational 
Research (IISER), Kolkata for the scientific discussion and the support 
when the maximum work of this study has been carried out. Also, SKB 
acknowledges the support by the NSFC (Grant No. 11750110422, 11433006, 
11790301, and 11790305). 
\bibliographystyle{spr-mp-sola}

\end{article} 

\end{document}